\documentclass[11pt,a4paper]{article}

\usepackage[margin=1in]{geometry}
\usepackage[T1]{fontenc}
\usepackage{microtype}
\usepackage{indentfirst}
\usepackage{amsmath,amssymb,mathtools,bm}

\usepackage{graphicx}
\usepackage{booktabs}
\usepackage{caption}
\usepackage[numbers,sort&compress]{natbib}

\usepackage[hidelinks]{hyperref}

\makeatletter
\renewcommand{\maketitle}{%
  \begingroup
  \centering

  \vspace*{-1.5em}

  \begin{minipage}{0.94\textwidth}
    \centering

    {\LARGE\bfseries
      \@title
      \par
    }

    \vspace{1.15em}

    {\large
      \@author
      \par
    }

    \vspace{0.85em}

    {\normalsize
      \@date
      \par
    }
  \end{minipage}

  \par
  \vspace{1.4em}
  \endgroup
}
\makeatother

\title{%
Exact Low-Dimensional Reduction Theory\\[0.12em]
for Populations of Stuart--Landau Oscillators%
}

\author{%
Kai Tokunaga\\[0.55em]
{\small
Department of Complexity Science and Engineering\\
The University of Tokyo, Kashiwa, Chiba 277-8561, Japan\\[0.35em]
\href{mailto:tokunaga.kai25a@c.k.u-tokyo.ac.jp}
{\nolinkurl{tokunaga.kai25a@c.k.u-tokyo.ac.jp}}%
}%
}

\date{5 August 2026}

\begin{document}

\maketitle

% ============================================================
% Abstract
% ============================================================
\begin{abstract}
This paper develops an exact low-dimensional reduction theory for populations of Stuart–Landau oscillators. The theory addresses two cases: (i) coupling enters only through the coefficients of the Stuart--Landau equations, and (ii) coupling also enters through terms outside those coefficients. Under suitable assumptions, the two cases can be reduced exactly to three- and seven-dimensional systems, respectively. The class of additional coupling terms for which the reduction applies is sufficiently general to allow a broad range of collective dynamics, such as clustering. In particular, the present reduction is shown to capture dynamics in which amplitude plays an essential role and which cannot be described by populations of phase oscillators or their low-dimensional reductions. The proposed framework is also shown to be a powerful tool for capturing complex nonequilibrium dynamics such as chaos.
\end{abstract}

% ============================================================
% Main text
% ============================================================
\section{Introduction}

Populations of coupled oscillators have long attracted attention because of their broad applications in biology, chemistry, engineering, physics, and other fields \cite{kuramoto1984chemical,winfree2001,Pikovsky_Rosenblum_Kurths_2001,strogatz2003sync}. Even when each oscillator obeys a simple ODE, a large population of coupled oscillators can form a highly complex dynamical system. Watanabe and Strogatz \cite{Watanabe1993} (WS) made a major breakthrough in the analysis of such systems. They showed that a population of identical phase oscillators with sinusoidal coupling can be reduced to a three-dimensional ODE system with $N-3$ constants of motion. In a subsequent major development, Ott and Antonsen \cite{Ott2008} (OA) showed that, in the thermodynamic limit, a population of phase oscillators with sinusoidal coupling can be reduced to a dynamical system on a low-dimensional invariant manifold. In particular, when the heterogeneity follows a Cauchy distribution, the system can be reduced to a two-dimensional ODE system. Since then, there have been many further developments in both exact \cite{Cestnik2022,Pietras2016,Skardal2011,Gong2019,Martens2009} and approximate \cite{Tyulkina2018,Vlasov_2016,Buendia2025,tokunaga2026} low-dimensional reduction theories for populations of phase oscillators.

By contrast, low-dimensional reductions for more general coupled oscillators with amplitude dynamics have remained less developed than those for phase oscillators. In this context, Cestnik and Martens \cite{Cestnik2024} made an important advance. By generalizing WS theory to populations of complex Riccati equations, they obtained an exact reduction to a six-dimensional system. Augustsson, Martens, and Cestnik \cite{Augustsson2026} also derived a low-dimensional reduction for populations of Mth-order quasilinear ODEs to a system of M+1 ODEs of order M. However, in both cases, each unit in the population, on its own, lacks an isolated stable periodic orbit and therefore cannot sustain stable oscillations. Units that exhibit stable self-sustained oscillations—namely, limit-cycle oscillators—are common in real-world systems, and populations of such oscillators have played a central role in synchronization theory \cite{kuramoto1984chemical,winfree2001,Pikovsky_Rosenblum_Kurths_2001,strogatz2003sync}. The Stuart–Landau oscillator \cite{landau1944,Stuart_1960} is a prototypical limit-cycle oscillator known as the normal form of a Hopf bifurcation \cite{kuramoto1984chemical}. This paper shows that populations of Stuart–Landau oscillators admit an exact low-dimensional reduction.

\section{Exact Low-Dimensional Reduction Theory
for Populations of Stuart--Landau Oscillators}
\subsection{Populations with Coupling through the Coefficients}

First, consider a population of \(N\) Stuart--Landau oscillators coupled through the coefficients:
\begin{equation}
\dot{z}_i = (\mu(t)+i\omega(t))z_i-[a(t)+ib(t)]z_i|z_i|^2 \quad(i=1,\dots,N).
\end{equation}
Here, \(\mu(t),\omega(t),a(t)\), and \(b(t)\) may be taken to be general real-valued functions of time \(t\) that include mean-field coupling, common forcing, and other effects. The important point is that these functions are independent of the index \(i\). Writing this equation in polar coordinates, \(z_i=r_i e^{i\theta_i}\), gives
\begin{equation}
\begin{split}
&\dot{r}_i=\mu(t)r_i-a(t)r_i^3 \\
&\dot{\theta}_i=\omega(t)-b(t)r_i^2.
\end{split}
\end{equation}
Since the amplitude equation is a Bernoulli equation, the change of variables \(u_i=1/r_i^2\) reduces it to the following linear ODE:
\begin{equation}
\dot{u}_i=-2\mu(t)u_i+2a(t).
\end{equation}
Define \(P(t),Q(t)\in\mathbb{R}\) by
\begin{equation}
\begin{split}
&\dot{P}=-2\mu(t)P\\
&\dot{Q}=-2\mu(t)Q+2a(t).
\end{split}
\end{equation}
The solution of Eq.~(3) can then be written using constants \(\xi_i\in\mathbb{R}\) as
\begin{equation}
u_i=P\xi_i+Q.
\end{equation}
It follows from this form of the solution that there are \(N-2\) independent constants of motion.

Next, consider the phase dynamics. The transformation
\(\phi_i=\theta_i-\frac{b(t)}{a(t)}\log r_i\) gives
\begin{equation}
\dot{\phi}_i
=
\omega(t)-\frac{b(t)}{a(t)}\mu(t)
-\left(\frac{d}{dt}\frac{b(t)}{a(t)}\right)\log r_i.
\end{equation}
Suppose that \(\frac{b(t)}{a(t)}=c\), that is, the nonisochronicity is a constant independent of time. The right-hand side is then independent of \(i\), and the \(N-1\) phase differences
\(\phi_i-\phi_1\) (\(i=2,\ldots,N\)) are constants of motion. Writing
\(\phi_i=\Phi_i+\Psi\), where \(\Phi_i\) are phase constants, gives
\begin{equation}
\dot{\Psi}=\omega(t)-c\mu(t).
\end{equation}
Together with \(P\) and \(Q\), this yields a real three-dimensional reduced system. However, because the phase differences are constants of motion, this model does not generally exhibit collective phase phenomena such as synchronization or clustering. Moreover, from
\(\theta_i=\Phi_i+c\log r_i+\Psi\), it follows that \(z_i\) is related to \(P,Q\), and \(\Psi\) by
\begin{equation}
z_i=e^{i(\Phi_i+\Psi)}(P\xi_i+Q)^{-\frac{1+ic}{2}}.
\end{equation}
The variables \(P\) and \(Q\) obtained here may diverge when an oscillator approaches the origin. Such divergences can be removed by a change of coordinates (see the Supplemental Material).

To summarize, the following population of Stuart--Landau oscillators with coupling through the coefficients and time-independent nonisochronicity,
\begin{equation}
\dot{z}_i = (\mu(t)+i\omega(t))z_i-a(t)[1+ic]z_i|z_i|^2,
\end{equation}
is reduced through the transformation in Eq.~(8) to \(2N-3\) constants of motion and the following three-dimensional system:
\begin{equation}
\begin{split}
&\dot{P}=-2\mu(t)P,\\
&\dot{Q}=-2\mu(t)Q+2a(t),\\
&\dot{\Psi}=\omega(t)-c\mu(t).
\end{split}
\end{equation}

\subsection{Populations with Coupling beyond the Coefficients}

To consider models that allow more complex collective behavior, including synchronization and clustering, consider polynomial coupling terms in addition to coupling through the coefficients:
\begin{equation}
\dot{z}_i = (\mu(t)+i\omega(t))z_i-a(t)[1+ic]z_i|z_i|^2
+\sum_{p,q\geq 0}H_{p,q}z_i^p\bar{z}_i^q.
\end{equation}
Here, \(H_{p,q}\) are general complex-valued functions of time that can represent effects such as mean-field coupling and common forcing. Writing the equation in terms of amplitude and phase gives
\begin{equation}
\begin{split}
&\dot{r}_i
=
\mu(t)r_i-a(t)r_i^3
+\sum_{p,q\geq 0}r_i^{p+q}
\operatorname{Re}\left(H_{p,q}e^{i(p-q-1)\theta_i}\right)\\
&\dot{\theta}_i
=
\omega(t)-a(t)cr_i^2
+\sum_{p,q\geq 0}r_i^{p+q-1}
\operatorname{Im}\left(H_{p,q}e^{i(p-q-1)\theta_i}\right).\\
\end{split}
\end{equation}
Unlike the case of coupling only through the coefficients, the amplitude equation is not decoupled from the phase. The phase equation also remains coupled to the amplitude even after applying \(\phi_i=\theta_i-c\log r_i\) as above. However, when \(c=0\), that is, in the isochronous case, a closed equation for the phase can be obtained by restricting the form of the coupling. Requiring the amplitude equation to be of Bernoulli type further restricts the coupling, resulting in the following ODE:
\begin{equation}
\dot{z}_i
=
(\mu(t)+i\omega(t))z_i-a(t)z_i|z_i|^2
+H(t)\bar{z}_i
+\operatorname{Re}\left(G(t)z_i^2\right)z_i.
\end{equation}
Writing the equation in terms of amplitude and phase gives
\begin{equation}
\dot{r}_i
=
\left[\mu(t)+\operatorname{Re}\left(H(t)e^{-i2\theta_i}\right)\right]r_i
-
\left[a(t)-\operatorname{Re}\left(G(t)e^{i2\theta_i}\right)\right]r_i^3
\end{equation}
\begin{equation}
\dot{\theta}_i
=
\omega(t)+\operatorname{Im}\left(H(t)e^{-i2\theta_i}\right).
\end{equation}
The phase equation thus contains second-harmonic coupling. Clustering phenomena can therefore be expected. On the other hand, studies of phase oscillators have shown that even with second-harmonic coupling, three-body interactions can produce asymmetric clustering \cite{Komarov2015,Gong2019,Moyal2024,Jain2025}. When this asymmetry is strong, the magnitude of the first-order parameter, which measures the degree of synchronization, becomes large. Thus, depending on the form of the interaction, the present system may also exhibit behavior resembling synchronization.

First, consider the phase equation. It has the form of a globally coupled system of identical phase oscillators with second-harmonic coupling. The WS theory applies to equations of this form \cite{Gong2019}. Specifically, let \(\zeta_i\) be constants on the unit circle. Using the following M\"obius transformation,
\begin{equation}
e^{i2\theta_i}
=
\frac{\alpha+e^{i\chi}\zeta_i}
{1+\bar{\alpha}e^{i\chi}\zeta_i},
\end{equation}
the phase dynamics can be reduced to the following three-dimensional ODE system, with \(N-3\) independent constants of motion:
\begin{equation}
\dot{\alpha}
=
2i\omega\alpha+H-\bar{H}\alpha^2
\end{equation}
\begin{equation}
\dot{\chi}
=
2\omega+2\operatorname{Im}\left(H(t)\bar{\alpha}\right).
\end{equation}
Here, \(|\alpha(t)|\leq 1\), and \(\chi(t)\) is a rotation angle.

Next, consider the amplitude equation. The change of variables \(u_i=1/r_i^2\) reduces it to the following linear ODE:
\begin{equation}
\dot{u}_i
=
-2\left[\mu(t)+\operatorname{Re}\left(H(t)e^{-i2\theta_i}\right)\right]u_i
+2\left[a(t)-\operatorname{Re}\left(G(t)e^{i2\theta_i}\right)\right].
\end{equation}
Unlike in the case of coupling through the coefficients, the coefficients of this linear ODE now depend on \(i\). To remove the term proportional to \(u_i\), we use the following identity:
\begin{equation}
\frac{d}{dt}
\left[
\log\left|1+\bar{\alpha}e^{i\chi}\zeta_i\right|^2
\right]
=
2\operatorname{Re}
\left[
H(t)\left(e^{-i2\theta_i}-\bar{\alpha}\right)
\right].
\end{equation}
Therefore, defining
\(v_i=\left|1+\bar{\alpha}(t)e^{i\chi(t)}\zeta_i\right|^2u_i\), we obtain
\begin{equation}
\begin{split}
\dot{v}_i
={}&
-2\left(\mu(t)+\operatorname{Re}(H\bar{\alpha})\right)v_i
+2\left|1+\bar{\alpha}e^{i\chi}\zeta_i\right|^2
\left[a(t)-\operatorname{Re}\left(G(t)e^{i2\theta_i}\right)\right]
\\
={}&
-2\left(\mu(t)+\operatorname{Re}(H\bar{\alpha})\right)v_i
\\
&+
2\left[
a(t)(1+|\alpha|^2)
-2\operatorname{Re}\left(G(t)\alpha\right)
\right]
\\
&+
2\operatorname{Re}
\left\{
\zeta_i e^{i\chi}
\left[
2a(t)\bar{\alpha}
-G(t)
-\bar{G}(t)\bar{\alpha}^2
\right]
\right\}.
\end{split}
\end{equation}
Define \(R(t),S(t)\in\mathbb{R}\) and \(T(t)\in\mathbb{C}\) by
\begin{equation}
\begin{split}
&\dot{R}
=
-2\left(\mu(t)+\operatorname{Re}(H\bar{\alpha})\right)R\\
&\dot{S}
=
-2\left(\mu(t)+\operatorname{Re}(H\bar{\alpha})\right)S
+2\left[
a(t)(1+|\alpha|^2)
-2\operatorname{Re}\left(G(t)\alpha\right)
\right]\\
&\dot{T}
=
-2\left(\mu(t)+\operatorname{Re}(H\bar{\alpha})\right)T
+e^{i\chi}
\left[
2a(t)\bar{\alpha}
-G(t)
-\bar{G}(t)\bar{\alpha}^2
\right].
\end{split}
\end{equation}
The solution of Eq.~(21) can then be written in terms of constants \(C_i\in\mathbb{R}\) as
\begin{equation}
v_i=C_iR+S+2\operatorname{Re}(T\zeta_i).
\end{equation}
This form shows that there are \(N-4\) independent constants of motion. Thus, the full system has \(2N-7\) independent constants of motion and is reduced to a seven-dimensional ODE system. The original variables \(z_i\) are reconstructed as
\begin{equation}
z_i
=
\sigma_i
\frac{
e^{i\frac{\chi}{2}}\zeta_i^{\frac{1}{2}}
+\alpha e^{-i\frac{\chi}{2}}\bar{\zeta}_i^{\frac{1}{2}}
}{
\left[
C_iR+S+2\operatorname{Re}(T\zeta_i)
\right]^{1/2}
}.
\end{equation}
Here, \(\sigma_i\in\{-1,1\}\) arises because the M\"obius transformation is applied to the second harmonic and is uniquely determined by the initial conditions (see the Supplemental Material). The variables \(R,S\), and \(T\) obtained here can diverge if an oscillator approaches the origin, but these divergences can be removed by a change of coordinates (see the Supplemental Material).

To summarize, the following isochronous population of Stuart--Landau oscillators with coupling beyond the coefficients,
\begin{equation}
\dot{z}_i
=
(\mu(t)+i\omega(t))z_i
-a(t)z_i|z_i|^2
+H(t)\bar{z}_i
+\operatorname{Re}\left(G(t)z_i^2\right)z_i,
\end{equation}
has \(2N-7\) independent constants of motion and is reduced, through the transformation in Eq.~(24), to the following seven-dimensional ODE system:
\begin{equation}
\begin{split}
&\dot{\alpha}
=
2i\omega\alpha+H-\bar{H}\alpha^2\\
&\dot{\chi}
=
2\omega+2\operatorname{Im}\left(H(t)\bar{\alpha}\right)\\
&\dot{R}
=
-2\left(\mu(t)+\operatorname{Re}(H\bar{\alpha})\right)R\\
&\dot{S}
=
-2\left(\mu(t)+\operatorname{Re}(H\bar{\alpha})\right)S
+2\left[
a(t)(1+|\alpha|^2)
-2\operatorname{Re}\left(G(t)\alpha\right)
\right]\\
&\dot{T}
=
-2\left(\mu(t)+\operatorname{Re}(H\bar{\alpha})\right)T
+e^{i\chi}
\left[
2a(t)\bar{\alpha}
-G(t)
-\bar{G}(t)\bar{\alpha}^2
\right].\\
\end{split}
\end{equation}

\section{Example}

\subsection{A Mean-Field-Coupled Population of Stuart--Landau Oscillators Subject to Common Forcing}

In this section, the reduction theory developed above is used to analyze a mean-field-coupled population of Stuart--Landau oscillators subject to common forcing. Here, \(H\) is taken to be
\(H=Ke^{i\tau}Z_2+Fe^{i2\Omega t}\), where
\(Z_2=\frac{1}{N}\sum_{j=1}^N z_j^2\) is the mean field and
\(Fe^{i2\Omega t}\) represents common forcing at frequency \(2\Omega\). The parameters \(\mu\), \(\omega\), and \(a\) are taken to be constants, and \(G=0\). The condition \(K<a\) is also assumed to prevent unphysical solutions with diverging amplitudes. A periodic forcing term of the form \(\bar{z}_iFe^{i2\Omega t}\) typically gives rise to a \(2:1\) resonance \cite{COULLET1992,Burke2008,YOCHELIS2004}.

In the rotating frame 
\(w_i=e^{-i\Omega t}z_i\), the system becomes autonomous:
\begin{equation}
\dot{w}_i
=
(\mu+i\Delta)w_i-aw_i|w_i|^2+H(t)\bar{w}_i
\end{equation}
\begin{equation}
H(t)
=
Ke^{i\tau}W_2+F.
\end{equation}
Here, \(\Delta=\omega-\Omega\), \(W_2=\frac{1}{N}\sum_{j=1}^N w_j^2\). The origin \(w_1=\cdots=w_N=0\) is clearly a fixed point of this system. Physically, the origin corresponds to the quiescent state (QS), in which no oscillation amplitude is present. Writing \(w_i=x_i+iy_i\), the linearized equation about this fixed point is
\begin{equation}
\frac{d}{dt}
\begin{pmatrix}
x\\
y
\end{pmatrix}
=
\begin{pmatrix}
\mu+F & -\Delta\\
\Delta & \mu-F
\end{pmatrix}
\begin{pmatrix}
x\\
y
\end{pmatrix}.
\end{equation}
The eigenvalues are therefore
\begin{equation}
\lambda
=
\mu\pm\sqrt{F^2-\Delta^2}.
\end{equation}
Thus, even when \(\mu<0\), the origin becomes unstable if
\(F>\sqrt{\mu^2+\Delta^2}\). In other words, even inactive oscillators with \(\mu<0\) can be excited into an oscillatory state by the common forcing.

A complete-clustering (CC) state of the form
\(r_i=\rho\), \(e^{i\theta_i}=\sigma_i e^{i\Theta}\)
(\(i=1,\ldots,N\)), in which all amplitudes are equal, is a typical stable fixed point away from the origin. For \(\mu>0\), loss of stability of the CC fixed point leads to either oscillating-amplitude complete clustering (OACC), in which all amplitudes remain equal, or partial clustering (PC). OACC can arise through either a SNIC bifurcation or a Hopf bifurcation. For \(\mu<0\), the CC fixed point can disappear through a saddle-node bifurcation, in which case the QS at the origin and CC are bistable. The corresponding parameter conditions can be derived analytically, and their derivations are given in the Supplemental Material. Since all CC and OACC states observed in the simulations had equal amplitudes for all oscillators, this property will not be stated explicitly below.

\subsection{Numerical Simulations}
\begin{figure}[t]
    \centering
    \includegraphics[width=1.0\linewidth]{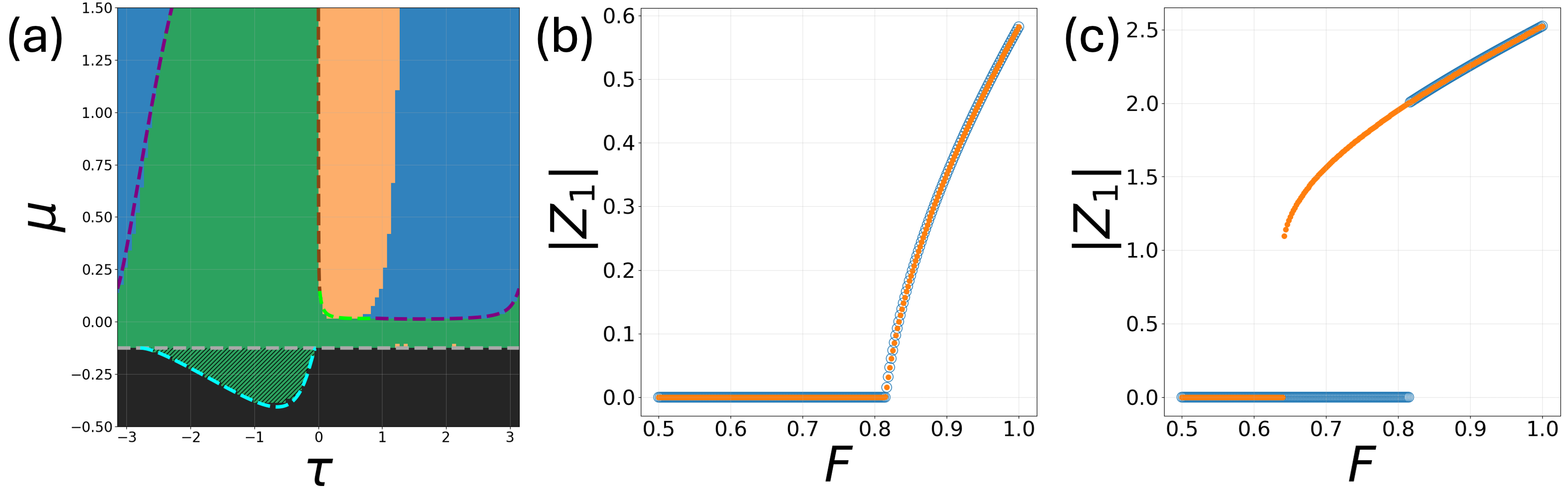}
    \caption{
        (a) Phase diagram. The analytically derived stability boundaries are shown in gray for the QS stability boundary, cyan for the boundary at which CC disappears in the bistable region, brown for the SNIC bifurcation from CC to OACC, yellow-green for the Hopf bifurcation from CC to OACC, and purple for the bifurcation from CC to PC. (b), (c) Transitions between QS and CC as \(F\) is decreased and increased. Blue open circles correspond to increasing \(F\), and orange dots correspond to decreasing \(F\).
    }
\end{figure}
We perform simulations using the seven-dimensional reduction. The initial amplitudes are drawn from the uniform distribution \(U(1,1.5)\), and the initial phases are drawn from the distribution
\(\rho(\theta)=\frac{1}{2\pi}\left[1+\frac{1}{2}\sin 2\theta\right]\).
For the reduced variables, we use the identity initial conditions
\(\alpha(0)=0\), \(\chi(0)=0\), \(R(0)=1\), \(S(0)=0\), and \(T(0)=0\).
With these choices,
\(z_i(0)=\frac{\sigma_i\zeta_i^{1/2}}{\sqrt{C_i}}\).
The seven-dimensional system can be closed by evaluating the mean field through the following transformation:
\begin{equation}
W_2
=
\frac{1}{N}\sum_{j=1}^N w_j^2
=
\frac{1}{N}\sum_{j=1}^N
\frac{
e^{i\chi}\zeta_j+2\alpha+\alpha^2e^{-i\chi}\bar{\zeta}_j
}{
C_jR+S+2\operatorname{Re}(T\zeta_j)
}.
\end{equation}

We simulate this seven-dimensional system and construct phase diagrams in the \(\tau\)--\(\mu\) plane. We consider the parameter sets
\(\Delta=0.79\), \(a=1\), \(F=0.8\), and \(K=0.8\) in Fig.~1(a), and
\(\Delta=0.1\), \(a=1\), \(F=0.25\), and \(K=0.3\) in Fig.~2(a).
The phase and amplitude order parameters are calculated from the corresponding transformations as follows:
\begin{equation}
A_2
=
\frac{1}{N}\sum_{j=1}^N e^{i2\theta_j}
=
\frac{1}{N}\sum_{j=1}^N
\frac{
\alpha+e^{i\chi}\zeta_j
}{
1+\bar{\alpha}e^{i\chi}\zeta_j
}
\end{equation}
\begin{equation}
B_2
=
\frac{1}{N}\sum_{j=1}^N r_j^2
=
\frac{1}{N}\sum_{j=1}^N
\frac{
\left|1+\bar{\alpha}e^{i\chi}\zeta_j\right|^2
}{
C_jR+S+2\operatorname{Re}(T\zeta_j)
}.
\end{equation}
The state with \(B_2=0\) is classified as the QS and is shown in black. A stationary state with \(|A_2|=1\) and \(B_2>0\) is classified as CC and is shown in green. The remaining states can be classified as OACC, for which \(|A_2|=1\) and \(B_2\) oscillates, or PC, for which \(0<|A_2|<1\). For the PC states, we additionally calculate the Lyapunov exponents. Chaotic regions in which the largest Lyapunov exponent is positive are shown in red. The phase diagrams also include curves representing the analytically derived stability boundaries of the fixed points.

Figures~1(a) and 2(a) show that the seven-dimensional reduction correctly reproduces the analytically obtained boundaries. As described above, inactive oscillators with \(\mu<0\) are excited by the periodic forcing and begin to oscillate, accompanied by a transition from QS to CC. For sufficiently large \(\mu\), increasing \(\tau\) causes a transition from CC to OACC. These amplitude-related dynamics cannot be captured by conventional theories for phase models. Further increasing \(\tau\) leads to PC. These features are common to both Figs.~1(a) and 2(a).

In Fig.~1(a), there is a green region with black hatching below the stability boundary of the QS. This is a region of bistability between the QS and CC. For the initial conditions used here, all trajectories approach CC, but the part with \(\mu\) below the analytically derived QS stability boundary is hatched. In Figs.~1(b) and 1(c), we set \(\Delta=0.79\), \(a=1\), and \(\mu=-0.2\), and vary \(F\) while monitoring \(|W_2|\) for \(K=0.2\) and \(K=0.8\), respectively. Since the reduced variables diverge as the QS is approached in the present coordinates, the integration is stopped once the state is identified as the QS. The figures show a continuous transition for \(K=0.2\), whereas an explosive transition occurs for \(K=0.8\). For \(K=0.8\), hysteresis caused by bistability is also observed. In the present model, an explosive transition does not occur when \(K=0\), and a sufficiently large \(K\) is required for it to occur. However, in models with nonisochronicity that are outside the scope of the present reduction, interactions between oscillators are not necessarily required for an explosive transition \cite{Burke2008,Edri2020}.

Unlike Fig.~1(a), Fig.~2(a) shows neither bistability between the QS and CC nor a transition from CC to OACC through a Hopf bifurcation. On the other hand, a chaotic region, which is absent in Fig.~1(a), appears near the boundary between PC and OACC. Figure~2(b) is a Feigenbaum diagram. For each value of \(\tau\), the values of \(|A_2|\) are recorded and plotted whenever the trajectory crosses the Poincar\'e section
\(\chi=0\pmod{2\pi}\), \(\dot{\chi}>0\).
This Feigenbaum diagram suggests that the observed chaos arises through a cascade of period-doubling bifurcations. Figure~2(c) further shows the corresponding points in the \((|A_2|,|B_2|)\) plane whenever the trajectory crosses the Poincar\'e section.
\begin{figure}[t]
    \centering
    \includegraphics[width=1.0\linewidth]{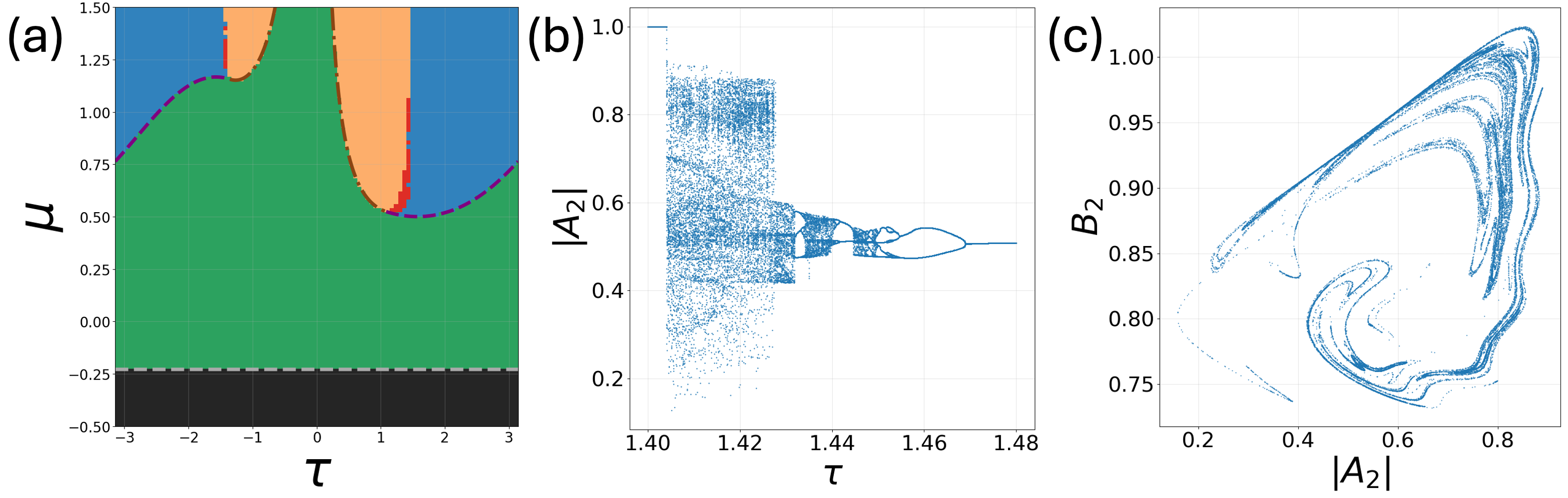}
    \caption{
        (a) Phase diagram. The colors of the stability boundaries are the same as in Fig.~1(a). (b) Feigenbaum diagram. (c) Values of \(|A_2|\) and \(B_2\) on the Poincar\'e section. In panels (b) and (c), \(\Delta=0.1\), \(a=1\), \(F=0.25\), \(K=0.3\), and \(\mu=0.85\). In panel (c), \(\tau=1.42\).
    }
\end{figure}
\section{Conclusion}
In this paper, exact low-dimensional reductions were derived for populations of Stuart–Landau oscillators. The theory applies to two classes of populations: (i) populations with coupling through the coefficients and constant nonisochronicity, and (ii) isochronous populations with coupling through the coefficients together with an additional coupling term of the form $H(t)\bar{z_i}+\operatorname{Re}(G(t)z_i^2)z_i$. It was shown that these two classes can be reduced to three- and seven-dimensional ODE systems, respectively. These results make it possible to apply low-dimensional reduction to populations of coupled oscillators whose individual units possess stable limit cycles on their own. This reduction theory is expected to play an important role in studying complex amplitude-mediated dynamics in populations of limit-cycle oscillators that phase-oscillator models cannot capture. Indeed, the reduced equations accurately capture several types of dynamics in which amplitude plays an essential role. These include oscillations induced in inactive oscillators by common forcing, the associated continuous or explosive transitions to complete clustering, and complete clustering with oscillating amplitudes. The reduction is also shown to be a powerful tool for capturing complex nonequilibrium dynamics such as chaos.

One important question suggested by the present results is whether, in the thermodynamic limit, a reduction to a low-dimensional invariant manifold is possible, as in the OA ansatz for populations of phase oscillators. For populations of complex Riccati equations, such a reduction has already been obtained for Cauchy-distributed heterogeneity \cite{Pazo2025}. It is therefore important to determine whether a similar reduction can be obtained for the populations of Stuart–Landau oscillators considered here. Another important direction is to examine the present reduction theory from a more algebraic perspective. WS theory is understood in terms of the action of the Möbius group \cite{Marvel2009}, and it would be interesting to determine the algebraic structure underlying the present reduction and to describe the theory from that perspective. Such an algebraic understanding may also help determine whether the forms of coupling considered here are the only ones that allow such a reduction for Stuart–Landau oscillators, or whether other forms are possible. In this regard, a framework based on Lie–Scheffers theory \cite{martens2026} that was recently presented in a preprint is expected to provide a powerful set of tools for addressing these questions.

\section*{Acknowledgments}

The author thanks Hiroshi Kori. This study was supported by Wings FMSP program of The Univ. of Tokyo.

% ============================================================
% References
% ============================================================
\bibliographystyle{unsrtnat}
\bibliography{Ref}

% ============================================================
% Supplemental Material
% ============================================================
\clearpage

\begin{center}
    {\Large\bfseries Supplemental Material}
\end{center}

\vspace{1em}

% Supplemental Materialの番号をS1, S2, ...にする
\setcounter{section}{0}
\setcounter{subsection}{0}
\setcounter{equation}{0}
\setcounter{figure}{0}
\setcounter{table}{0}

\renewcommand{\thesection}{S\arabic{section}}
\renewcommand{\thesubsection}{\thesection.\arabic{subsection}}
\renewcommand{\theequation}{S\arabic{equation}}
\renewcommand{\thefigure}{S\arabic{figure}}
\renewcommand{\thetable}{S\arabic{table}}

% Supplemental Material中のhyperref用アンカーを本体と区別する
\renewcommand{\theHsection}{supplement.\arabic{section}}
\renewcommand{\theHsubsection}
    {supplement.\arabic{section}.\arabic{subsection}}
\renewcommand{\theHequation}{supplement.\arabic{equation}}
\renewcommand{\theHfigure}{supplement.\arabic{figure}}
\renewcommand{\theHtable}{supplement.\arabic{table}}

\section{Reduced Systems in Bounded Coordinates}

\subsection{Coupling through the Coefficients}

The variables \(P\) and \(Q\) derived above may diverge when an oscillator approaches the origin. We therefore transform them to coordinates in which such divergences do not occur. Define \(\hat{P}\) and \(\hat{Q}\) by
\[
\hat{P}
=
\frac{P}{1+\sqrt{1+P^2+Q^2}},
\qquad
\hat{Q}
=
\frac{Q}{1+\sqrt{1+P^2+Q^2}}.
\]
Since \(\hat{P}^2+\hat{Q}^2<1\), these variables remain bounded. The inverse transformation is
\[
P
=
\frac{2\hat{P}}{1-(\hat{P}^2+\hat{Q}^2)},
\qquad
Q
=
\frac{2\hat{Q}}{1-(\hat{P}^2+\hat{Q}^2)},
\]
showing that \((P,Q)\) and \((\hat{P},\hat{Q})\) are in one-to-one correspondence. In terms of \(\hat{P}\) and \(\hat{Q}\), the transformation becomes
\begin{equation}
z_i
=
e^{i(\Phi_i+\Psi)}
\left[
\frac{1-(\hat{P}^2+\hat{Q}^2)}
{2(\hat{P}\xi_i+\hat{Q})}
\right]^{\frac{1+ic}{2}}.
\end{equation}
Furthermore, Eq.~(5) allows us to take \(\xi_i>0\) without loss of generality. Defining the single complex constant
\(
\nu_i
=
e^{i\Phi_i}\xi_i^{-\frac{1+ic}{2}},
\)
the transformation can be written as
\begin{equation}
z_i
=
\nu_i e^{i\Psi}
\left[
\frac{1-(\hat{P}^2+\hat{Q}^2)}
{2(\hat{P}+\hat{Q}|\nu_i|^2)}
\right]^{\frac{1+ic}{2}}.
\end{equation}
When \(z_i(0)=0\), setting \(\nu_i=0\) allows this transformation to be defined even for an initial condition at the origin, where the phase is undefined. The reduced three-dimensional system is
\begin{equation}
\begin{split}
&\dot{\hat{P}}
=
-\frac{2(1-\hat{P}^2-\hat{Q}^2)}
{1+\hat{P}^2+\hat{Q}^2}
\left[\mu(t)+a(t)\hat{Q}\right]\hat{P}\\
&\dot{\hat{Q}}
=
-\frac{2(1-\hat{P}^2-\hat{Q}^2)}
{1+\hat{P}^2+\hat{Q}^2}
\left[\mu(t)+a(t)\hat{Q}\right]\hat{Q}
+a(t)(1-\hat{P}^2-\hat{Q}^2)\\
&\dot{\Psi}
=
\omega(t)-c\mu(t).
\end{split}
\end{equation}

\subsection{Coupling beyond the Coefficients}

The variables \(R\), \(S\), and \(T\) derived above may diverge when an oscillator approaches the origin. We therefore transform them to coordinates in which such divergences do not occur. Define
\[
\hat{R}
=
\frac{R}{1+\sqrt{1+R^2+S^2+|T|^2}},
\,
\hat{S}
=
\frac{S}{1+\sqrt{1+R^2+S^2+|T|^2}},
\,
\hat{T}
=
\frac{T}{1+\sqrt{1+R^2+S^2+|T|^2}}.
\]
Since
\[
\hat{R}^2+\hat{S}^2+|\hat{T}|^2<1
\]
holds at all times, these variables remain bounded. The inverse transformation is
\[
R
=
\frac{2\hat{R}}
{1-(\hat{R}^2+\hat{S}^2+|\hat{T}|^2)},
\,
S
=
\frac{2\hat{S}}
{1-(\hat{R}^2+\hat{S}^2+|\hat{T}|^2)},
\,
T
=
\frac{2\hat{T}}
{1-(\hat{R}^2+\hat{S}^2+|\hat{T}|^2)},
\]
showing that the two sets of coordinates are in one-to-one correspondence. In terms of \(\hat{R}\), \(\hat{S}\), and \(\hat{T}\), the transformation becomes
\begin{equation}
z_i
=
\sigma_i
\sqrt{
\frac{1-(\hat{R}^2+\hat{S}^2+|\hat{T}|^2)}{2}
}
\frac{
e^{i\frac{\chi}{2}}\zeta_i^{\frac{1}{2}}
+\alpha e^{-i\frac{\chi}{2}}\bar{\zeta}_i^{\frac{1}{2}}
}{
\left[
C_i\hat{R}
+\hat{S}
+2\operatorname{Re}(\hat{T}\zeta_i)
\right]^{1/2}
}.
\end{equation}
Furthermore, Eq.~(23) allows us to take \(C_i>0\) without loss of generality. Combining the constants into the single complex constant
\(
\eta_i
=
\frac{\sigma_i\zeta_i^{\frac{1}{2}}}{\sqrt{C_i}},
\)
the transformation can be written as
\begin{equation}
z_i
=
\sqrt{
\frac{1-(\hat{R}^2+\hat{S}^2+|\hat{T}|^2)}{2}
}
\frac{
e^{i\frac{\chi}{2}}\eta_i
+\alpha e^{-i\frac{\chi}{2}}\bar{\eta}_i
}{
\left[
\hat{R}
+\hat{S}|\eta_i|^2
+2\operatorname{Re}(\hat{T}\eta_i^2)
\right]^{1/2}
}.
\end{equation}
When \(z_i(0)=0\), setting \(\eta_i=0\) allows this transformation to be defined even for an initial condition at the origin, where the phase is undefined. The reduced seven-dimensional system is
\begin{equation}
\begin{split}
&\dot{\alpha}
=
2i\omega\alpha+H-\bar{H}\alpha^2\\
&\dot{\chi}
=
2\omega+2\operatorname{Im}\left(H(t)\bar{\alpha}\right)\\
&\dot{\hat{R}}
=
-\Gamma(t)\hat{R}\\
&\dot{\hat{S}}
=
-\Gamma(t)\hat{S}
+
\left[
1-(\hat{R}^2+\hat{S}^2+|\hat{T}|^2)
\right]
\left[
a(t)(1+|\alpha|^2)
-2\operatorname{Re}\left(G(t)\alpha\right)
\right]\\
&\dot{\hat{T}}
=
-\Gamma(t)\hat{T}
+
\frac{
1-(\hat{R}^2+\hat{S}^2+|\hat{T}|^2)
}{2}
e^{i\chi}
\left[
2a(t)\bar{\alpha}
-G(t)
-\bar{G}(t)\bar{\alpha}^2
\right].\\
\end{split}
\end{equation}
Here,
\begin{equation}
\begin{split}
&\Gamma(t)
=
\frac{
1-(\hat{R}^2+\hat{S}^2+|\hat{T}|^2)
}{
1+\hat{R}^2+\hat{S}^2+|\hat{T}|^2
}
\Bigg[
2\left(
\mu(t)+\operatorname{Re}(H\bar{\alpha})
\right)
\\
&\quad
+2\hat{S}
\left[
a(t)(1+|\alpha|^2)
-2\operatorname{Re}\left(G(t)\alpha\right)
\right]
+\operatorname{Re}
\left\{
\overline{\hat{T}}e^{i\chi}
\left[
2a(t)\bar{\alpha}
-G(t)
-\bar{G}(t)\bar{\alpha}^2
\right]
\right\}
\Bigg].
\end{split}
\end{equation}

\section{Constancy of \(\sigma_i\) under Time Evolution}

Since
\begin{equation}
e^{i2\theta_i}
=
\frac{\alpha+e^{i\chi}\zeta_i}
{1+\bar{\alpha}e^{i\chi}\zeta_i}
=
\frac{
\left(
e^{i\frac{\chi}{2}}\zeta_i^{\frac{1}{2}}
+\alpha e^{-i\frac{\chi}{2}}\bar{\zeta}_i^{\frac{1}{2}}
\right)^2
}{
\left|
e^{i\frac{\chi}{2}}\zeta_i^{\frac{1}{2}}
+\alpha e^{-i\frac{\chi}{2}}\bar{\zeta}_i^{\frac{1}{2}}
\right|^2
},
\end{equation}
we have
\begin{equation}
\sigma_i
=
e^{i\theta_i}
\frac{
\left|
e^{i\frac{\chi}{2}}\zeta_i^{\frac{1}{2}}
+\alpha e^{-i\frac{\chi}{2}}\bar{\zeta}_i^{\frac{1}{2}}
\right|
}{
e^{i\frac{\chi}{2}}\zeta_i^{\frac{1}{2}}
+\alpha e^{-i\frac{\chi}{2}}\bar{\zeta}_i^{\frac{1}{2}}
}.
\end{equation}
When \(|\alpha|<1\), the right-hand side is continuous in time. Since \(\sigma_i\in\{-1,1\}\), it must therefore remain constant in time.

\section{Stability Boundaries of the Complete-Clustering Fixed Point}

For the complete-clustering (CC) state considered here, in which all amplitudes are equal, the dynamics lie on a two-dimensional invariant manifold, referred to as the complete-clustering manifold. The dynamics on this manifold satisfy the following two-dimensional system:
\begin{equation}
\begin{split}
&\dot{\rho}
=
\left[
\mu+F\cos 2\Theta+(K\cos\tau-a)\rho^2
\right]\rho\\
&\dot{\Theta}
=
\Delta+K\rho^2\sin\tau-F\sin 2\Theta.
\end{split}
\end{equation}

When CC is a fixed point, it is also a fixed point of this two-dimensional system. Let the fixed point be \((\rho^*,\Theta^*)\). Eliminating \(\Theta^*\) gives
\begin{equation}
g(\rho^{*\,2})
=
-F^2
+
\left[
\mu+(K\cos\tau-a)\rho^{*\,2}
\right]^2
+
\left[
\Delta+K\rho^{*\,2}\sin\tau
\right]^2
=
0.
\end{equation}

Linearizing Eqs.~(S10) around \((\rho^*,\Theta^*)\) gives the following Jacobian matrix:
\begin{equation}
J
=
2
\begin{pmatrix}
(K\cos\tau-a)\rho^{*\,2}
&
-\Delta\rho^*-K\rho^{*\,3}\sin\tau
\\
K\rho^*\sin\tau
&
\mu+(K\cos\tau-a)\rho^{*\,2}
\end{pmatrix}.
\end{equation}

Its determinant and trace are
\begin{equation}
\det J
=
4\left[
(K\cos\tau-a)^2+K^2\sin^2\tau
\right]\rho^{*\,4}
+
4\left[
\mu(K\cos\tau-a)+\Delta K\sin\tau
\right]\rho^{*\,2}
\end{equation}
\begin{equation}
\operatorname{tr}J
=
2\mu+4(K\cos\tau-a)\rho^{*\,2}.
\end{equation}

\subsection{Saddle-Node Bifurcation within the Complete-Clustering Manifold}

Since
\(\det J=2\rho^{*\,2}g'(\rho^{*\,2})\),
the condition \(\det J=0\) is equivalent to the double-root condition and gives the saddle-node bifurcation condition. At the saddle-node bifurcation, \(\det J=0\) gives
\(
\rho^{*\,2}
=
-\frac{
\mu(K\cos\tau-a)+\Delta K\sin\tau
}{
K^2+a^2-2Ka\cos\tau
}.
\)
Eliminating \(\rho^{*\,2}\) from Eq.~(S11) and requiring
\(\rho^{*\,2}>0\) gives the following conditions for a saddle-node bifurcation within the complete-clustering manifold:
\begin{equation}
F^2(K^2+a^2-2Ka\cos\tau)
-
\left[
\mu K\sin\tau-\Delta(K\cos\tau-a)
\right]^2
=
0
\end{equation}
\begin{equation}
\mu(K\cos\tau-a)+\Delta K\sin\tau
<
0.
\end{equation}

First, consider the case \(\mu<0\). The divergence of the vector field on the complete-clustering manifold is
\(
2\mu+4(K\cos\tau-a)|w_i|^2<0,
\)
so no periodic orbit exists. Rewriting Eq.~(S15) gives
\begin{equation}
F^2
=
\mu^2+\Delta^2
-
\frac{
\left[
\Delta K\sin\tau+\mu(K\cos\tau-a)
\right]^2
}{
K^2+a^2-2Ka\cos\tau
}
<
\mu^2+\Delta^2.
\end{equation}
Therefore, CC disappears through this saddle-node bifurcation before the QS loses stability. This gives bistability between CC and the QS. Furthermore, when \(K=0\), the expression in Eq.~(S16) becomes positive, and the inequality cannot be satisfied. Thus, interactions between oscillators are necessary for this phenomenon.

Next, consider the case \(\mu>0\). In this case, a SNIC bifurcation is expected, in which a limit cycle is created through a saddle-node bifurcation. Strictly speaking, a SNIC bifurcation is a global bifurcation and therefore cannot be identified from the present local analysis alone. However, numerical simulations confirm that a limit cycle is created after the saddle-node bifurcation. In addition to the saddle-node bifurcation conditions above, the trace \(\operatorname{tr}J\) must be negative for the limit cycle created by the SNIC bifurcation to be stable. This gives the additional condition
\begin{equation}
\mu
\left[
(K\cos\tau-a)^2-K^2\sin^2\tau
\right]
+
2\Delta K\sin\tau(K\cos\tau-a)
>
0.
\end{equation}

\subsection{Hopf Bifurcation within the Complete-Clustering Manifold}

A Hopf bifurcation occurs when
\(\operatorname{tr}J=0\) and \(\det J>0\). The condition
\(\operatorname{tr}J=0\) gives
\(
\rho^{*\,2}
=
\frac{\mu}{2(a-K\cos\tau)}
\)
at the Hopf bifurcation. Using Eq.~(S11) to eliminate
\(\rho^{*\,2}\), and requiring \(\rho^{*\,2}>0\) and
\(\det J>0\), gives the following conditions for a Hopf bifurcation within the complete-clustering manifold:
\begin{equation}
F^2
-
\frac{\mu^2}{4}
-
\left[
\Delta+
\frac{\mu K\sin\tau}{2(a-K\cos\tau)}
\right]^2
=
0
\end{equation}
\begin{equation}
\mu>0
\end{equation}
\begin{equation}
\mu
\left[
(K\cos\tau-a)^2-K^2\sin^2\tau
\right]
+
2\Delta K\sin\tau(K\cos\tau-a)
<
0.
\end{equation}

\subsection{Stability Transverse to the Complete-Clustering Manifold}

The saddle-node and Hopf bifurcations within the complete-clustering manifold are relevant to the phase diagram only when the complete-clustering manifold is stable. Partial clustering appears when the complete-clustering manifold itself loses stability. We therefore consider perturbations transverse to this manifold. Write a transverse perturbation as
\(
w_i=\sigma_i w^*+\delta w_i.
\)
For a transverse perturbation, we can impose
\(
\frac{1}{N}\sum_{i=1}^N\sigma_i\delta w_i=0.
\)
The linearized equation for the transverse perturbation is then
\begin{equation}
\dot{\delta w}_i
=
\left(
\mu+i\Delta-2a\rho^{*\,2}
\right)\delta w_i
-
(\mu+i\Delta)e^{2i\Theta^*}\overline{\delta w_i}.
\end{equation}
Here, the fixed-point condition has been used in the second term. Writing \(w_i=r_i e^{i\theta_i}\), the perturbation can be expressed as
\(
\delta w_i
=
\sigma_i e^{i\Theta^*}
\left(
\delta r_i+i\rho^*\delta\theta_i
\right).
\)
The linearized equation for \(\delta r_i\) and \(\delta\theta_i\) is therefore
\begin{equation}
\frac{d}{dt}
\begin{pmatrix}
\delta r_i\\
\delta\theta_i
\end{pmatrix}
=
2
\begin{pmatrix}
-a\rho^{*\,2} & -\Delta\rho^*\\
0 & \mu-a\rho^{*\,2}
\end{pmatrix}
\begin{pmatrix}
\delta r_i\\
\delta\theta_i
\end{pmatrix}.
\end{equation}

The eigenvalue in the amplitude direction is
\(
\lambda_r=-2a\rho^{*\,2}<0,
\)
and is therefore always stable. Thus, a pure amplitude mode does not become unstable on its own. The eigenvalue in the phase direction is
\(
\lambda_\theta=2(\mu-a\rho^{*\,2}),
\)
and is stable when
\(
\lambda_\theta=2(\mu-a\rho^{*\,2})<0.
\)
It is therefore always stable when \(\mu<0\). The SNIC and Hopf bifurcations within the complete-clustering manifold are relevant to the phase diagram when
\(\lambda_\theta<0\). This condition gives
\(
\mu(K-a\cos\tau)+\Delta a\sin\tau<0
\)
for the SNIC bifurcation, and
\(
K\cos\tau-\frac{a}{2}>0
\)
for the Hopf bifurcation.

The condition
\(
\lambda_\theta
=
2(\mu-a\rho^{*\,2})
=
0
\)
gives the bifurcation at which CC loses stability and PC appears. Eliminating \(\rho^{*\,2}\) using Eq.~(S11), and requiring
\(\rho^{*\,2}>0\), gives the following conditions for the bifurcation to PC:
\begin{equation}
a^2F^2
-
\mu^2K^2\cos^2\tau
-
(a\Delta+\mu K\sin\tau)^2
=
0
\end{equation}
\begin{equation}
\mu>0.
\end{equation}
For the bifurcation to PC to occur first, no bifurcation within the complete-clustering manifold must have occurred. Therefore,
\(\det J>0\) and \(\operatorname{tr}J<0\) must hold, giving the additional conditions
\begin{equation}
K\cos\tau-\frac{a}{2}<0
\end{equation}
\begin{equation}
\mu(K-a\cos\tau)+\Delta a\sin\tau>0.
\end{equation}

\subsection{Summary}

The loss of stability of the CC fixed point can therefore be classified as follows:
\begin{itemize}

\item Disappearance of CC through a saddle-node bifurcation while CC and the QS are bistable:
\begin{flalign*}
&\mu<0&\\
&F^2(K^2+a^2-2Ka\cos\tau)
-
\left[
\mu K\sin\tau-\Delta(K\cos\tau-a)
\right]^2
=
0&\\
&\mu(K\cos\tau-a)+\Delta K\sin\tau<0.
\end{flalign*}
\item SNIC bifurcation to OACC:
\begin{flalign*}
&\mu>0&\\
&F^2(K^2+a^2-2Ka\cos\tau)
-
\left[
\mu K\sin\tau-\Delta(K\cos\tau-a)
\right]^2
=
0\\
&\mu(K\cos\tau-a)+\Delta K\sin\tau<0&\\
&\mu
\left[
(K\cos\tau-a)^2-K^2\sin^2\tau
\right]
+
2\Delta K\sin\tau(K\cos\tau-a)
>
0&\\
&\mu(K-a\cos\tau)+\Delta a\sin\tau<0.
\end{flalign*}
\item Hopf bifurcation to OACC:
\begin{flalign*}
&\mu>0&\\
&F^2
-
\frac{\mu^2}{4}
-
\left[
\Delta+
\frac{\mu K\sin\tau}{2(a-K\cos\tau)}
\right]^2
=
0&\\
&\mu
\left[
(K\cos\tau-a)^2-K^2\sin^2\tau
\right]
+
2\Delta K\sin\tau(K\cos\tau-a)
<
0&\\
&K\cos\tau-\frac{a}{2}>0.
\end{flalign*}
\item Bifurcation to PC:
\begin{flalign*}
&\mu>0&\\
&a^2F^2
-
\mu^2K^2\cos^2\tau
-
(a\Delta+\mu K\sin\tau)^2
=
0&\\
&K\cos\tau-\frac{a}{2}<0&\\
&\mu(K-a\cos\tau)+\Delta a\sin\tau>0&.
\end{flalign*}

\end{itemize}

\end{document}